\documentclass[pra,aps,groupedaddress,twocolumn,floatfix,nofootinbib]{revtex4}
\usepackage{graphicx}

\newcommand{\beq}{\begin{equation}}
\newcommand{\eeq}{\end{equation}}
\newcommand{\beqa}{\begin{eqnarray}}
\newcommand{\eeqa}{\end{eqnarray}}

\newcommand{\braket}[2]{\mbox{$ \langle #1 | #2 \rangle $}}

\newcommand{\ket}[1]{\mbox{$ | #1 \rangle $}}
\newcommand{\bra}[1]{\mbox{$ \langle #1 | $}}

\def\half{\frac{1}{2}}
\def\opone{\leavevmode\hbox{\small1\normalsize\kern-.33em1}}
\def\ptr{p_{tr}}

\begin{document}

\title{The Elegant Joint Quantum Measurement and some conjectures about N-locality in the Triangle and other Configurations}
\author{Nicolas Gisin}

\date{\small \today}
\begin{abstract}In order to study N-locality without inputs in long lines and in configurations with loops, e.g. the triangle, we introduce a natural joint measurement on two qubits different from the usual Bell state measurement. The resulting quantum probability $p(a_1,a_2,...,a_N)$ has interesting features. In particular the probability that all results are equal is that large, while respecting full symmetry, that it seems highly implausible that one could reproduce it with any N-local model, though - unfortunately - I have not been unable to prove it.
\end{abstract}
\maketitle

\section{Introduction}
3-locality in the triangle configuration \cite{triangle} has by now been studied by a rather large group of scientists over several years, see Fig. 1. It is fair to admit that not much has been found. Actually, the only quantum example is due to Tobbias Fritz \cite{TFritz12}. However, this nice example is essentially the well known CHSH Bell inequality folded into a triangle: the usual suspects Alice and Bob form the base of the triangle and the inputs are provided by a referee located at the top (3rd vertex - Charlie) of the triangle.

The main goal of most research on the triangle is to find a quantum example without inputs, i.e. an example with three 2-partite quantum states and one quantum measurement per party, that is provably not 3-local. Recall that the set of 3-local probabilities $p(a,b,c)$ is not convex, hence the usual numerical tools based on geometric polytopes don't work: the analog of Bell inequalities for 3-locality are non-linear \cite{triangle}. Most research concentrated on the case of binary outcomes. A first pretty trivial conjecture, motivated by the lack of success, is that such an example doesn't exist.

Considering 3 independent singlets shared pairwise between A, B and C it is natural to expect that quantum examples use joint measurement on 2 qubits with 4 outcomes each. The first idea uses the well known Bell State Measurements (BSM). However, it is not difficult to show that the corresponding joint correlation $p(a,b,c)$ has a 3-local model (in this note we use the two terms probability and correlation for $p(a,b,c)$). Hence, one should look for other joint measurements. The fact is that there are not many "elegant" joint measurements on 2 qubits, whatever "elegant" means. For example, if the eigenstates of the joint measurement are all maximally entangled, then the BSM is unique. Hence, let's consider joint measurements for which all eigenstates have the same degree of partial entanglement. Moreover, we like the partial states of these eigenstates to point to the tetrahedron on the Poincar\'e (or Bloch) sphere, i.e. to display maximal symmetry.

\begin{figure}
\includegraphics[width=6cm]{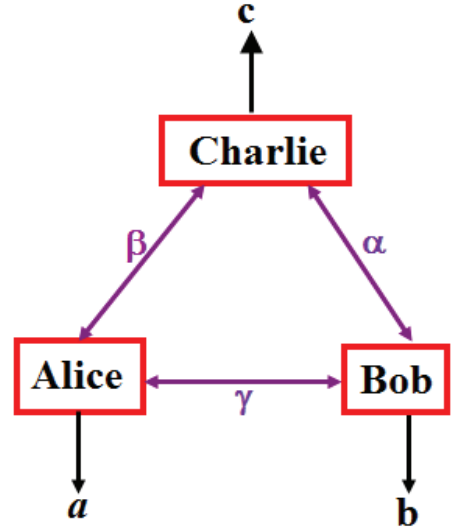}
\vspace{-10pt}
\caption{\it The triangle configuration for 3 parties. Each pair of parties shares either a quantum state and performs quantum measurements - quantum scenario, or shares independent random variables $\alpha$, $\beta$ and $\gamma$ and outputs a function of the random variables to which they have access. Notice that the three random variables are only used locally, hence the terminology 3-local scenario. The ``Quantum Grail'' is to find a quantum scenario leading to a probability $p(a,b,c)$ which can't be reproduced by any 3-local scenario.}
\end{figure}

In the next section we show that the above mentioned constrains define a unique Elegant Joint Measurement (EJM). Then, in section 3 we compute the corresponding correlation and studied some of its elementary property. From these we conjecture, in section 4, that this correlation is not 3-local, though - unfortunately - without proof. Next we present the use of our EJM for n-locality in a line and N-vertex polygons.

\section{The Elegant Joint Measurement on 2 qubits}
Denote the 4 vertices of the tetrahedron as follows:
\beqa
\vec m_1&=& (1,1,1)/\sqrt{3} \label{m1} \\
\vec m_2&=& (1,-1,-1)/\sqrt{3} \label{m2} \\
\vec m_3&=& (-1,1,-1)/\sqrt{3} \label{m3} \\
\vec m_4&=& (-1,-1,1)/\sqrt{3} \label{m4} 
\eeqa

Using cylindrical coordinates, $\vec m_j=(\sqrt{1-\eta_j^2}\cos{\phi_j}, \sqrt{1-\eta_j^2}\sin{\phi_j},\eta_j)$, one obtains the natural correspondence with qubit states:
\beq\label{ketm}
\ket{\vec m_j}=\sqrt{\frac{1-\eta_j}{2}}e^{i\phi_j/2}\ket{0}+\sqrt{\frac{1+\eta_j}{2}}e^{-i\phi_j/2} \ket{1}
\eeq
Note that $\vec m_j=\bra{\vec m_j}\vec\sigma\ket{\vec m_j}$, as expected (with $\vec\sigma$ the 3 Pauli matrices). \\

Massar and Popescu \cite{MassarPopescu} introduced joint measurements inspired by parallel spins appropriately superposed with the singlet, $\psi^-=\ket{01-10}/\sqrt{2}$:
\beqa
\Psi_1=\frac{\sqrt{3}}{2}\ket{\vec m_1,\vec m_1}+\half\psi^- \\
\Psi_2=\frac{\sqrt{3}}{2}\ket{\vec m_2,\vec m_2}-\half\psi^- \\
\Psi_3=\frac{\sqrt{3}}{2}\ket{\vec m_3,\vec m_3}-\half\psi^- \\
\Psi_4=\frac{\sqrt{3}}{2}\ket{\vec m_4,\vec m_4}+\half\psi^- \\
\eeqa
Note that the probability amplitudes and phases are imposed by the requirement that the $\Psi_j$'s should be normalized and mutually orthogonal. Importantly, the asymmetric phases are unavoidable. Furthermore, although the 4 $\Psi_j$'s have the same degree of entanglement, the following ugly happens: the Bloch vectors of the partial states on Alice and on Bob sides are distorted and squashed towards the equator: the 4 vectors $\bra{\Psi_j}\vec\sigma\otimes\opone\ket{\Psi_j}$ point into different directions than the vectors (\ref{m1}-\ref{m4}). 

Consequently, and inspired by \cite{antiParallelSpins}, consider the following 2-qubit basis constructed on anti-parallel spins:
\beqa\label{Psij}
\Phi_j=\sqrt{\frac{3}{2}}\ket{\vec m_j, -\vec m_j} + i\frac{\sqrt{3}-1}{2}\psi^-
\eeqa
where $\ket{-\vec m}$ is orthogonal to $\ket{\vec m}$, it has the same form as (\ref{ketm}) but with $\eta\rightarrow -\eta$ and $\phi\rightarrow\phi+\pi$. 

In order to check that the $\Phi_j$ are normalised and mutually orthogonal one should use $\braket{\vec m, -\vec m}{\psi^-}=i/\sqrt{2}$ for all $\vec m$ and $\braket{\vec m_j, -\vec m_j}{\vec m_k, -\vec m_k}=1/3$ for all $j\neq k$.

Finally, for all $j$ one has $\bra{\Phi_j}\vec\sigma\otimes\opone\ket{\Phi_j}=\half\vec m_j$ and $\bra{\Phi_j}\opone\otimes\vec\sigma\ket{\Phi_j}=-\half\vec m_j$, hence $|\bra{\Phi_j}\vec\sigma\otimes\opone\ket{\Phi_j}|=\frac{\sqrt{3}}{2}$.\\

We name the 2-qubit measurement with eigenstates (\ref{Psij}) the Elegant Joint Measurement (EJM). We believe it is unique with all 4 eigenstates having identical degrees of partial entanglement and with all partial states of all eigenstates parallel or anti-parallel to the vertices of the tetrahedron.

\section{Quantum Correlation from singlets and the EJM in the triangle configuration}
Consider 3 independent singlets in the triangle configuration and assume that Alice, Bob and Charlie each perform the EJM on their 2 (independent) qubits, see Fig. 1. Denote the resulting correlation $\ptr(a,b,c)$, where $a,b,c=1,2,3,4$. By symmetry $p(a,b,c)$ is fully characterized by 3 numbers corresponding to the cases $a=b=c$, $a=b\ne c$ and $a\ne b\ne c\ne a$. A not too complex computation gives:
\beqa
\ptr(a=k,b=k,c=k)&=&\frac{25}{256}~ for~ k=1,2,3,4\\
\ptr(a=k,b=k,c=m)&=&\frac{1}{256}~ for~ k\ne m \\
\ptr(a=k,b=n,c=m)&=&\frac{5}{256}~ for~ k\ne n\ne m\ne k\nonumber\\ \label{kdiffndiffm}
\eeqa
The normalization holds: $4\cdot\frac{25}{256}+36\cdot\frac{1}{256}+24\cdot\frac{5}{256}=1$.

As expected $\ptr(a)=\ptr(b)=\ptr(c)=\frac{1}{4}$. More interesting is the probabilities that two parties get identical results:
\beqa
\ptr(a&=&k,b=k)=\nonumber\\
&=&\ptr(a=b=c=k)+\ptr(a=b=k,c\ne k)\nonumber\\
&=&\frac{25+3\cdot 1}{256}=\frac{7}{64}
\eeqa
Hence, all pairs of parties are correlated, e.g. $\ptr(a|b)\ne\frac{1}{4}$. In worlds, given an outcome $b=k$ for Bob, Alice's outcome has a large chance to take the same value: $\ptr(a=k|b=k)=\frac{\ptr(a=k,b=k)}{\ptr(b=k)}=\frac{7}{16}$. Accordingly:
\beq
\ptr(a=b)=\sum_k \ptr(b=k)p(a=k|b=k)=\frac{7}{16}
\eeq

The strength of the 3-party correlation is even more impressive:
\beq
\ptr(a=k|b=c=k)=\frac{\ptr(a=b=c=k)}{\ptr(b=c=k)}=\frac{25}{28}
\eeq
Hence $\ptr(a=b=c)=\frac{4\cdot25}{256}=\frac{25}{64}$.

\section{Is $\ptr(a,b,c)$ 3-local?}
In this section we consider the question whether the quantum probability $\ptr(a,b,c)$ is 3-local, i.e. whether it can be reproduced by a 3-local model.
In such a 3-local model of $\ptr(a,b,c)$ the Alice-Bob correlation could only be due to their shared randomness $\gamma$. Similarly, the correlation between Bob and Charlie is necessarily due to $\alpha$ and the Alice-Charlie correlation due to $\beta$. Accordingly, each local variable $\alpha$, $\beta$ and $\gamma$ contains a 4-dit, equally distributed among the values 1,2,3,4, and with a relatively high probability both Alice and Bob output the 4-dit contained in $\gamma$, and similarly for the other pairs of parties. Admittedly, this is only an argument, not a proof.

Accordingly, let's consider the following natural type of 3-local models. Let $\gamma=(\gamma_1,\gamma_2)$, where $\gamma_1=1,2,3,4$ with equal probability and $\gamma_2=0,1$ with $prob(\gamma_2=1)=q$. The idea is that whenever $\gamma_2=1$, then Alice and Bob results are given by $\gamma_1$, hence Alice and Bob get perfectly correlated. More explicitly, Alice's output function reads:
\beq\label{abetagamma}
a(\beta,\gamma)=\left\{
\begin{array}{c}
	\hspace{0.5cm}\gamma_1\hspace{5mm} if \hspace{2mm}\beta_2=0\hspace{2mm} and \hspace{2mm}\gamma_2=1\\
  \hspace{0.5cm}\beta_1\hspace{0.5cm}if~\beta_2=1\hspace{2mm}and\hspace{2mm}\gamma_2=0 \\
  \beta_1|\gamma_1\hspace{2mm}if\hspace{2mm}\beta_2=\gamma_2\hspace{15mm}
\end{array}
\right.
\eeq
where $\beta_1|\gamma_1$ indicates that $a(\beta,\gamma)$ equals $\beta_1$ or $\gamma_1$ with equal probability $\half$.

Table I indicates all possible outputs (where $\bar q\equiv(1-q)=prob(\alpha_2=0)=prob(\beta_2=0)=prob(\gamma_2=0))$.
\begin{table}[h]
	\centering
		\begin{tabular}
			{c|c|c|c|c|c|c|c|c}
			\large
			$\alpha_2$&$\beta_2$&$\gamma_2$&a&b&c&P&prob(a=b)&prob(a=b=c)\\
			\hline
			0&0&0&$\beta_1|\gamma_1$&$\alpha_1|\gamma_1$&$\alpha_1|\beta_1$&$\bar q^3$&7/16&13/64 \\
			0&0&1&$\gamma_1$&$\gamma_1$&$\alpha_1|\beta_1$&$\bar q^2q$&1&1/4 \\
			0&1&0&$\beta_1$&$\alpha_1|\gamma_1$&$\beta_1$&$\bar q^2q$&1/4&1/4 \\
			0&1&1&$\beta_1|\gamma_1$&$\gamma_1$&$\beta_1$&$\bar q q^2$&5/8&1/4 \\
			1&0&0&$\beta_1|\gamma_1$&$\alpha_1$&$\alpha_1$&$\bar q^2q$&1/4&1/4 \\
			1&0&1&$\gamma_1$&$\alpha_1|\gamma_1$&$\alpha_1$&$\bar q q^2$&5/8&1/4 \\
			1&1&0&$\beta_1$&$\alpha_1$&$\alpha_1|\beta_1$&$\bar q q^2$&1/4&1/4 \\
			1&1&1&$\beta_1|\gamma_1$&$\alpha_1|\gamma_1$&$\alpha_1|\beta_1$&$q^3$&7/16&13/64 \\
		\end{tabular}
\caption{\it The 8 lines correspond to the 8 possible combinations of values of $\alpha_2$, $\beta_2$ and $\gamma_2$ (first 3 columns). The next 3 columns indicate Alice, Bob and Charlie's outputs. The 7th column indicates the probability of the corresponding line and the last two columns the probability that $a=b$ and $a=b=c$, respectively.}
\end{table}

Averaging the probabilities that $a=b=c$ over the 8 combinations of values of $\alpha_2$, $\beta_2$ and $\gamma_2$, i.e. over the 8 lines of Table 1, gives:
\beqa
p_{3loc}(a=b=c)&=&\frac{13}{64}(\bar q^3+q^3)+\frac{3}{4}(\bar q^2q+\bar q q^2) \nonumber\\
&=&\frac{13+9q-9q^2}{64}
\eeqa
Hence, the maximal 3-partite correlation of our 3-local model is achieved for $q=\half$ and reads:
\beq
\max_q~p_{3loc}(a=b=c)=\frac{61}{256}
\eeq
This is much smaller than the value obtained in the quantum case with the Elegant Joint Measurement.

The above is not a proof, but leads us to conjecture that the quantum probability $\ptr(a,b,c)$ is not 3-local. Indeed, $\gamma$ has to correlated A an B, i.e. contribute to the probability that $a=b$, and $\beta$ contribute to $\ptr(a=c)$ and $\gamma$ contribute to $\ptr(a=b)$. But then the three independent variables $\alpha$, $\beta$ and $\gamma$ can't do the job for the 3-partite correlation $a=b=c$.

Note that if the outcomes are groups 2 by 2, such that outcomes are binary, then a 3-local model similar to (\ref{abetagamma}) can reproduce the quantum correlation. But, again, with 4 outcomes per party this seems impossible.

\subsection{A natural but asymmetric 3-local model}
There is another 3-local model that we need to consider, directly inspired by the quantum singlet states shared by each pair of parties. Assumes that the three local variables $\alpha$, $\beta$ and $\gamma$ each take values (0,1) and (1,0) with 50\% probabilities, where $\alpha$ send its first bit to Bob and second bit to Charlie, and similarly for $\beta$ and $\gamma$. Clearly, this 3-local model assumes binary local variables, i.e. bits,  but we like to keep the notation (0,1) and (1,0) for the two values.

The outcomes are then determined by the two bits that each party receives from the local variables it shares with his two neighbors. We like to maximize the probability $p(a=b=c)$. All output functions that maximize $p(a=b=c)$ are equivalent. On possible choice  is:
\beqa
(0,0)\Rightarrow a=2,~b=4,~c=3~~ \\
(0,1)\Rightarrow a=1,~b=1,~c=1~~ \\
(1,0)\Rightarrow a=3,~b=2,~c=4~~ \\
(1,1)\Rightarrow a=4,~b=3,~c=2~~
\eeqa
Note that in this 3-local model $\gamma$ imposes that both Alice and Bob can only output one out of two values. Which of the two values happen depens on the second local variables. This provides intuition why this 3-local model achieves $p(a=b=c)=\half$, i.e. an even larger value than the quantum probabilities with the EJM. Moreover $p(a=b)=\half$, hence $p(a=b=c|b=c)=1$. However, this model does not respect the symmetries of the quantum scenario. In particular 20 out of 24 cases $p(a=k,b=n,c=m)$ with $k\ne n\ne m\ne k$ take values 0 (recall that in the quantum scenario all 24 probabilities take value $\frac{5}{256}$, see eq. (\ref{kdiffndiffm})).

This simple 3-local model shows that in order to prove the non-3-locality of $\ptr(a,b,c)$ it is not sufficient to consider $p(a=b=c)$, but one has to consider also the cases $a\ne b\ne c$.

\section{The EJM in a line and arbitrary polygon}
Consider a polygon with N vertices, i.e. N parties and N independent singlets shared by each pair of parties connected by an edge, see Fig. 2. Assume all parties perform the EJM. It is not too difficult to compute the probability that $n<N$ neighbors get the same result, $p_{line}(a_1=a_2=...=a_n)$.
The subscript ``line'' indicates that this probability is the same as in the case of n parties in a line, connected by $n-1$ singlets, with 2 open-ended singlets at each end, see Fig. 3. It is also not too difficult to find the probability that all N parties in the polygon configuration get the same result, $p_{polygon}(a_1=a_2=...=a_N)$.

\begin{figure}
\includegraphics[width=6cm]{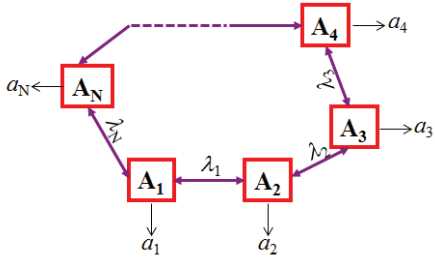}
\vspace{-10pt}
\caption{\it The N-vertex polygon configuration for N parties. The $\lambda_j$'s may represent quantum states - quantum scenario - or independent random variables - N-local scenario.}
\end{figure}

\begin{figure}
\includegraphics[width=9cm]{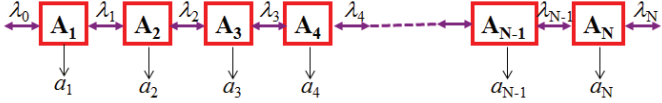}
\vspace{-10pt}
\caption{\it The N-vertex configuration for N parties on a line. Note the $\lambda_0$ and $\lambda_N$ on the far left and right ends, corresponding to an ``open line''.}
\end{figure}

For these computation best is to start with the following state constructed not with the vectors (\ref{m1}-\ref{m4}), but with the vectors $\pm\vec e_z$:
\beqa
\Phi&=&\sqrt{\frac{3}{2}}\ket{0,1}-\frac{\sqrt{3}-1}{2}\psi^-\\
&=&\frac{\sqrt{3}+1}{2\sqrt{2}}\ket{0,1}+\frac{\sqrt{3}-1}{2\sqrt{2}}\ket{1,0}
\eeqa
where we used $\ket{0}^\perp=-i\ket{1}$ and we dropped an irrelevant global phase. We get:
\beq\label{nEqual}
p_{line}(a_1=a_2=...=a_n)=\frac{(\sqrt{3}+1)^{2n}+(\sqrt{3}-1)^{2n}}{2^{4n-1}}
\eeq
\beq\label{NEqual}
p_{polygon}(a_1=a_2=...=a_N)=\frac{\big( (-\sqrt{3}-1)^N + (\sqrt{3}-1)^N\big)^2}{4^{2N-1}}
\eeq

Table II provides these probabilities for N up to ten.
\begin{table}[h]
	\centering
		\begin{tabular}
			{c|c|c|c}
			\large
			$N$&line&polygon&$p(a_1=...=a_N|pa_1=...a_{N-1})$\\
			\hline
			1&1& &\\
			2&7/16&1&1\\
			3&13/64&25/64&25/28\\
			4&97$\cdot 2^{-10}$&49/256&49/52\\
			5&$181\cdot 2^{-12}$&$361\cdot 2^{-12}$&361/388\\
			6&$1351\cdot 2^{-16}$&$169\cdot 2^{-12}$&169/181\\
			7&$2521\cdot 2^{-18}$&$5'041\cdot 2^{-18}$&5041/5404\\
			8&$18'817\cdot 2^{-22}$&$9'409\cdot 2^{-20}$&9409/10084\\
			9&$35'113\cdot 2^{-24}$&$70'225\cdot 2^{-24}$&70225/75268\\
			10&$262'087\cdot 2^{-28}$&$32'761\cdot 2^{-24}$&32761/35113\\
		\end{tabular}
\caption{\it Probabilities that all N results are equal for the (open) line and polygon configurations. The last column shows the probability that the Nth result equals all previous results, conditioned that the first N-1 results are all equal (polygon configuration).}
\end{table}

\begin{figure}
\includegraphics[width=6cm]{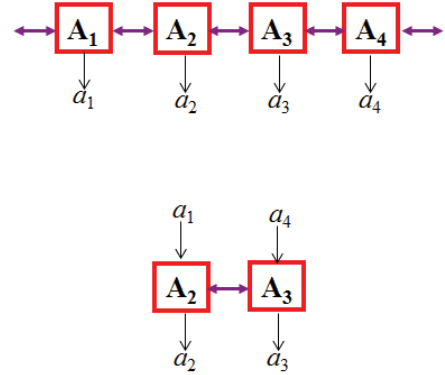}
\vspace{-10pt}
\caption{\it Relation between the open line with $N=4$ parties in the open line configuration (top) and standard Bell locality (bottom).}
\end{figure}

Let us first consider the line with N parties, more precisely the open line since parties $A_1$ and $A_N$ get half singlets and output thus random 4-dits $a_1$ and $a_N$, respectively, see Fig. 4 (top). These two random 4-dits can equally be considered as inputs for parties $A_2$ and $A_{N-1}$, see Fig. 4 (bottom). For N=4 we thus recover a standard 2-partite Bell scenario which, unfortunately, has local correlation (assuming all parties perform the EJM and that my linprog software is correct). For N=5 we recover standard bilocality \cite{bilocality} and for general N, we recover ($N-2$)-locality where only the first and last parties get inputs.

Let's now consider a polygon with N parties. The fact that the probability that all N outcomes are equal given that N-1 are equal is very large it tends asymptotically to:
\beqa
&& p_{polygon}(a_1=a_2=...=a_N|a_1=a_2=...=a_{N-1})\nonumber\\
&& \stackrel{N=\infty}{\longrightarrow} \frac{1}{8-4\sqrt{3}}\approx  93.3\% 
\eeqa
Hence, I conjecture that $p_{polygon}(a_1,a_2,...,a_N)$ is not N-local for all $N\ge3$.

\section{Conclusion}
N-locality in a loop (polygon) is a hard problem. But it is a fascinating one! Looking for a quantum example it is temping to consider that all neighbors share singlets and perform joint quantum measurements on their two qubits with 4 possible outcomes. Since the usual Bell state measurement doesn't lead to non-N-locality, it is natural to look for other joint measurements. Assuming all four eigenstates have the same degree of partial entanglement and all partial states display the natural tetrahedron symmetry, there seem to be only one such Elegant Joint Measurement. Consequently, we study quantum scenarios in polygons with N parties that all perform the EJM, in particular with N=3, i.e. the infamous triangle, and find that the quantum probability $p_{polygon}(a_1,...,a_N)$ displays strong correlations between any sets of n neighbors. The probability that all N outcome are equal, see eq (\ref{NEqual}), is that large that it seems impossible to obtain in any symmetric N-local model. Hence we conjecture that $p_{polygon}(a_1,...,a_N)$ is non-N-local for all $N\ge3$. We give some arguments in favour of our conjecture, though without proof.


\begin{thebibliography}{99}
\bibitem{triangle} C. Branciard, D. Rosset, N. Gisin, and S. Pironio, Phys. Rev. A {\bf 85}, 032119 (2012).
\bibitem{TFritz12} T. Fritz, New Journal of Physics {\bf 14}, 103001 (2012).
\bibitem{MassarPopescu} S. Massar and S. Popescu, Phys. Rev. Lett. {\bf 74}, 1259 (1995).
\bibitem{antiParallelSpins} N. Gisin and S. Popescu, Phys. Rev. Lett. {\bf83}, 432 (1999).
\bibitem{bilocality} C. Branciard, N. Gisin, and S. Pironio, Phys. Rev. Lett. {\bf 104}, 170401 (2010).
\bibitem{etc} etc, etc
\end{thebibliography}
\end{document}